\documentclass[11pt,fullpage]{article}

\usepackage[utf8]{inputenc} 
\usepackage[T1]{fontenc}    
\usepackage{hyperref}       
\usepackage{url}            
\usepackage{booktabs}       
\usepackage{amsfonts}       
\usepackage{nicefrac}       
\usepackage{microtype}      
\usepackage{amsmath,graphicx}

\usepackage{authblk}

\usepackage[utf8]{inputenc} 
\usepackage[T1]{fontenc}    
\usepackage{hyperref}       
\usepackage{url}            
\usepackage{booktabs}       
\usepackage{amsfonts}       
\usepackage{nicefrac}       
\usepackage{microtype}      
\usepackage{amsmath,graphicx}
\usepackage{amsmath}

\title{Exploring Generative Physics Models with Scientific Priors\\ in Inertial Confinement Fusion}

\author{Rushil Anirudh\footnote{Corresponding Author: anirudh1@llnl.gov}, Jayaraman J. Thiagarajan, Shusen Liu,\\  Peer-Timo Bremer, Brian K. Spears}
 \affil[]{Lawrence Livermore National Laboratory, Livermore, California.}
 \date{}
\begin{document}

\maketitle

\begin{abstract}
There is significant interest in using modern neural networks for scientific applications due to their effectiveness in modeling highly complex, non-linear problems in a data-driven fashion. However, a common challenge is to verify the scientific plausibility or validity of outputs predicted by a neural network. This work advocates the use of known scientific constraints as a lens into evaluating, exploring, and understanding such predictions for the problem of inertial confinement fusion.

 \let\thefootnote\relax \footnotetext{This work was performed under the auspices of the U.S. Department of Energy by Lawrence Livermore National Laboratory under Contract DE-AC52-07NA27344.}
\end{abstract}

\section{Introduction}

Modern neural networks are highly effective in modeling complex, multi-modal data and thus have raised significant interested in exploiting these capabilities for scientific applications.
In particular, the ability to directly ingest multi-modal, non-scalar data, i.e.\ images, energy spectra, etc., has proven to be a significant advantage over more traditional statistical approaches. 
Examples range from using neural networks to build advanced surrogate models of expensive simulations~\cite{kim2015time,peurifoy2018nanophotonic} to using generative models to cheaply sample complex data distributions~\cite{mustafa2019cosmogan,paganini2018calogan}.
One common challenge for such systems is to properly account for various invariants and constraints to guarantee physically meaningful results, i.e.\ positive energy, mass conservation, etc.
Existing approaches either integrate the physical laws, or rather the corresponding partial differential equations, directly into the training process~\cite{raissi2019physics,zhu2019physics} or add the constraints into the loss function~\cite{karpatne2017physics}.
However, this only works for known constraints that can be explicitly formulated as some differentiable equation in order to be integrated into the neural network training. 
In practice, not all constraints are known or can be formulated in this manner and explicitly enforcing some constraints while ignoring others is likely to bias the resulting system. 
Furthermore, constraints are often based on unrealistic assumptions, i.e. physical relationships under some idealized condition, which are not satisfied in the real data. 
Consequently, strictly enforcing such constrains may produce incorrect results.

Here we are interested in exploring, evaluating, and understanding the behavior of generative models~\cite{GANGoodfellow, kingma2013auto, tolstikhin2017wasserstein} for scientific datasets. 
Generative models have revolutionized our ability to sample from complex, high dimensional image distributions, but their application to scientific data is relatively new. 
In particular, it is not clear how to properly evaluate such models in this context.
By directly incorporating all known constraints into the loss function, evaluating the constraints post-hoc becomes a self-fulfilling prophecy with the compliance driven largely by the choice of weights in the loss function and a significant potential to over-correct the results. 
At the same time, most existing metrics are either designed for traditional computer vision problems like Inception scores \cite{salimans2016improved}, FID-scores \cite{heusel2017gans}, or they rely on other global metrics like manifold alignment~\cite{khrulkov2018geometry}, which may have little significance in the scientific context. 
Instead, we propose to use the constraints to evaluate a generative model and show how exploring the data distribution in latent space, i.e.\ the physics manifold, through the lense of the constraint can provide interesting insights. 
In particular, we use Inertial Confinement Fusion (ICF)~\cite{kritcher2014metrics} as a testbed problem, with multi-modal data generated from a 1D semi-analytic simulator~\cite{gaffney2014thermodynamic}. 

As will be discussed in more detail below, the simulator outputs four images as well as a number of scalars which create a complex high dimensional distribution.
The brightness integrals of the four images are related to the temperature (one of the scalars) and for an idealized plasma this relationship is a known prior. 
However, it is challenging to incorporate this term into the loss function.
More importantly the simulated system is not ``ideal'' and thus even the original simulator does not strictly obey the constraint. 
Instead, we consider the constraint to be a \emph{scientific prior} and explore how this prior is expressed in a Wasserstein auto-encoder  of the outputs. 
Interestingly, we find that the auto-encoder naturally organizes the data with respect to the prior with one of its latent dimension parametrizing the data according to the temperature-brightness relationship. 
Furthermore, the prior provides an intuitive lens through which one can explore the \emph{physics manifold} in the latent space. 
By scanning through various directions of the latent space one can observe that varying some latent dimensions quickly violates the prior and thus leaves the physics manifold, while others seem virtually invariant with respect to the prior. 
Finally, using the expected prior we show that despite the Gaussian prior imposed during the training the tails of the latent space distribution are far from Gaussian.

\section{Inertial Confinement Fusion}
\label{sec:methods}

Inertial confinement fusion (ICF) \cite{lindl1995development} experiments  use powerful lasers to heat and compress a millimeter-scale target filled with thermonuclear deuterium-tritium (DT) fuel.  The goal is to drive fusion reactions that self-heat the DT fuel leading to ignition and propagating burn.  The experiments employ a variety of diagnostic instruments to observe the fusion implosions.  These include exceptionally fast x-ray cameras and neutron cameras that create spectrally resolved (i.e. hypercolor) images, as well as a host of spectrometers and radiochemical diagnostics that produce key scalar features.  The fusion science teams use this multimodal set of observations as feedback to optimize the driving conditions of the experiment in a quest to improve the implosion performance.
 
Traditionally, ICF scientists use predictive forward models to produce simulated diagnostic output, which they later compare with experiments.  Recent work has used inference methods using deep neural networks as surrogates for the complicated radiation hydrodynamics codes that map laser and target conditions into output physics observations of images and scalars.  However, our physics application demands that the deep neural networks provide outputs that are consistent with physical laws, that constrain these outputs by establishing necessary functional relationships amongst them.  We formulate these scientific priors for all of the data modalities, providing us with individual priors amongst the scalars, priors amongst the images, and finally priors on relationships between images and scalars. As an example in this work, we can use notions of thermal equilibrium to relate predicted ion temperatures (as part of the scalars) to estimates of electron temperature formed by ratios of x-ray image brightnesses. As shown in the experiments, this manifests as a strong linear relationship between the predicted scalar, and the mean brightness of the x-ray images.  This constraint, and others allow us to judge the physical plausibility of DNN model predictions.  Those predictions that satisfy our ion/electron temperature requirements are admissible; those that violate them are ruled unphysical.
\section{Results}

\begin{figure*}[!thb]
	\centering
	\includegraphics[trim={0 0.6in 0 0},clip,width=0.9\linewidth]{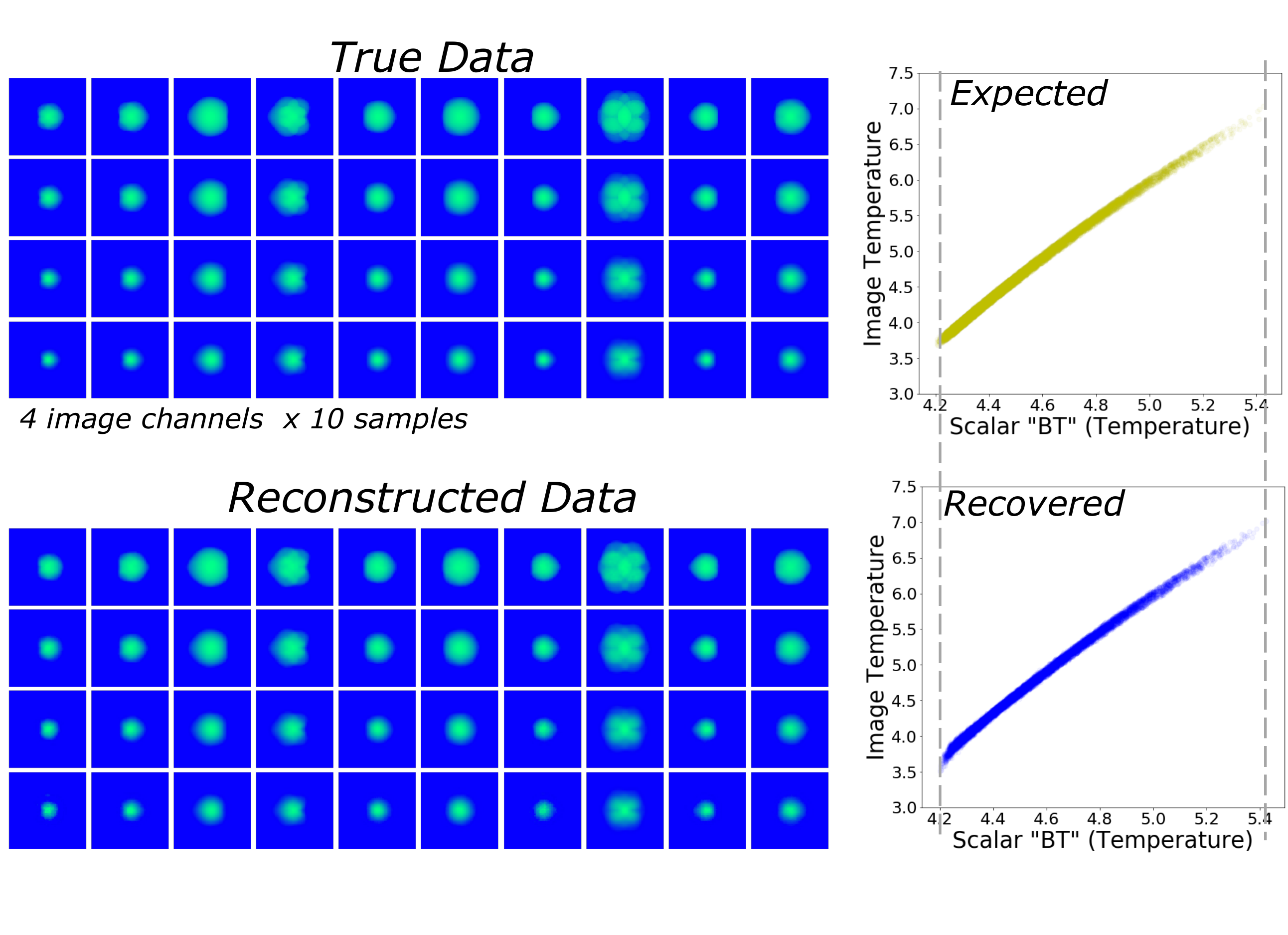}
	\caption{ Reconstructions from the trained autoencoder (left) and evaluating the scientific prior on the original and reconstructed data (right).}
	\label{fig:ae_recon}
\end{figure*}

\textbf{Dataset Description:} We use a 1D semi-analytic model simulator to produce a training dataset of size 100K, and a validation set of 10K samples. The data consists of 4 images of size $64\times64$, where each image corresponds to an x-ray measurement at pre-defined energy. We treat these 4 images as channels resulting in an image of $64\times64\times4$. Aside from images, we also have 15 unique scalar quantities that measure quantities such as yield, ion temperature, pressure, and other physical quantities. One of the ways in which the images and scalars are correlated is defined by the scientific prior defined in Section \ref{sec:methods}.

\textbf{Autoencoder:} We train a Wasserstein autoencoder \cite{tolstikhin2017wasserstein} on the images and scalars and embed it into a 10-dimensional latent space. We use convolutional layers for the images and fully connected layers for the scalars. These are stacked in the layer before the bottleneck and unstacked a layer after the bottleneck. We use a mean squared error (MSE) loss on both of them while giving the scalar reconstruction loss a weight of $1e^2$ to ensure accurate reconstruction. We use the Adam optimizer with a learning rate of $1e-4$, and a mini-batch size of 128 samples. Additionally, we use a discriminator network to guide the latent space distribution to be approximately Gaussian. We use an adversarial weight of $\lambda_{adv} = 0.05$ while training the autoencoder. A few test reconstructions and their corresponding true images across the 4 channels are shown in figure \ref{fig:ae_recon}. Along with the images, the autoencoder also produces the corresponding 15 scalars per sample, which are not shown here since they are fairly easy to recover almost perfectly (R2 score = 0.99). Besides each set of images, we also show the scientific prior in a plot (each point in the plot is a sample) that shows a clear linear relationship between the estimated image temperature, and the Scalar "BT" that represents measured temperature. In the bottom, we show the corresponding plot for the reconstructed data and it can be seen that the autoencoder does an excellent job of respecting the linear relationship, without explicitly being constraint to do so.

\begin{figure*}[!thb]
	\centering
	\includegraphics[trim={0 0 0 0},clip,width=0.9\linewidth]{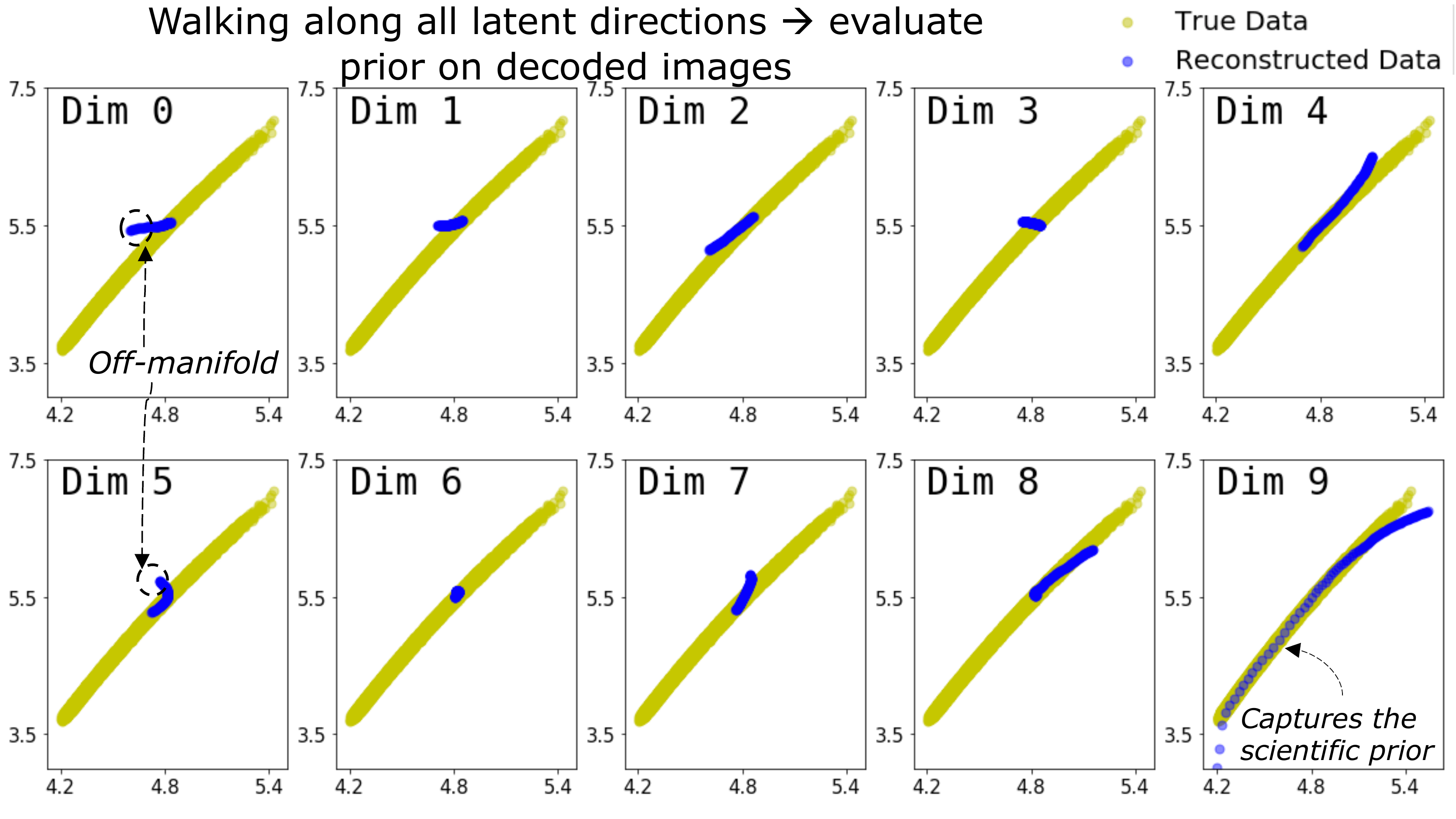}
	\caption{For a given sample in roughly the average position, we walk along each of the latent dimensions (while keeping other dimensions fixed) using a linear scan from its min to max value; this is followed by decoding each of those new latent vectors and computing the scientific prior for each decoded sample. We find that the last dimension in the latent space almost perfectly captures the scientific prior. Scientific priors allow us to visualize how  particular scientific property of interest varies as we navigate in the high dimensional latent space. }
	\label{fig:walking}
\end{figure*}

\textbf{Walking in the latent space:} Next in Figure~\ref{fig:walking}, we explore the latent manifold by walking along all the 10 dimensions using a linear scan from its min to max value. That is, for a given sample, we create a new artificial dataset given by $\mathbf{Z}^0$, where the $0^{th}$ dimension is replaced by a $100$ values between $z^0_{min} \dots z^0_{max}$, whereas $\{z^1, z^2, \dots, z^9\}$ are kept fixed. This is repeated for each of the 10 dimensions, as shown in Figure~\ref{fig:walking}. We observe a few key details with this experiment: 
\begin{itemize}
\item Most of the dimensions are \emph{insensitive} to the scientific prior, since the prior is a necessary but not sufficient condition it cannot capture all details about the physics manifold. 
\item Only one of the dimensions (dim 9) captures the scientific prior almost perfectly. This is important because this dimension can be used as a proxy to evaluate different aspects of physics.
\item The starting point of each of these `paths' is determined by which sample is chosen to perform this experiment, here we choose a sample that is approximately in the middle of the line. 
\item If we interpret this scientific prior as a 1D manifold projection of the high dimensional physics data, we can see that some directions in the latent space can lead to "off-manifold" samples. This is seen for dimensions 0 and 5. In some cases such as dimensions 4 and 9, the decoded samples eventually move off-manifold because we are increasingly sampling in the tail of the marginal distribution.
\end{itemize}

\begin{figure*}[!thb]
	\centering
	\includegraphics[trim={0 0 0 0},clip,width=0.9\linewidth]{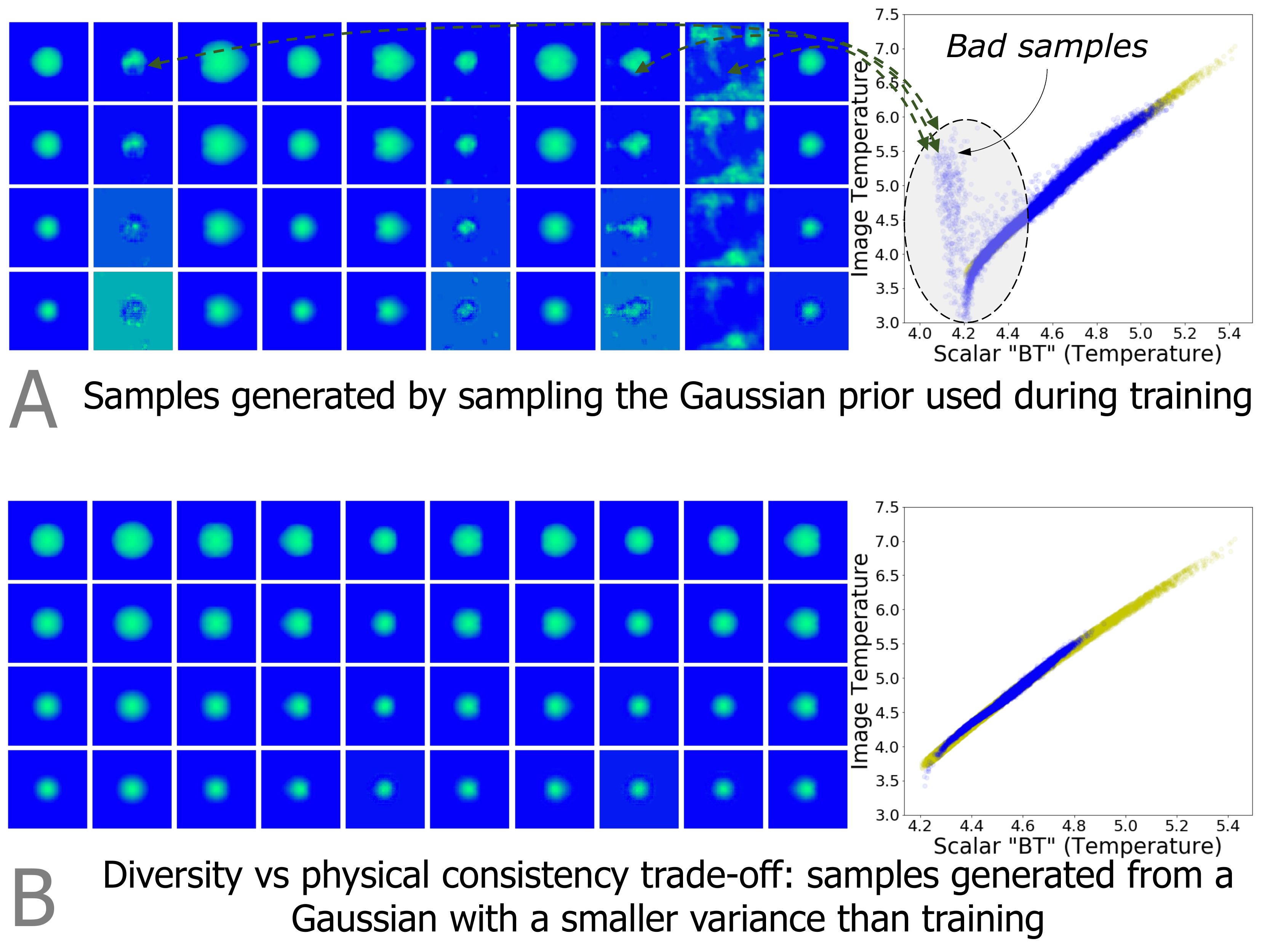}
	\caption{Generating new samples, and examining them through the the lens of the scientific prior.}
	\label{fig:sampling}
\end{figure*}

\textbf{Evaluating generative models with scientific priors} We also investigate how the scientific prior can help us better understand the quality of generative models. Since we impose a Gaussian prior on the latent space, we can sample from it after training to obtain new samples. Specifically, we sample from $z \sim \mathcal{N}(0.5, 0.5)$, and pass these samples through the decoder. Some of these samples are shown in Figure~\ref{fig:sampling}(A), where it is seen that the quality of samples is poor. We also show the corresponding scientific prior on the right side, where these bad samples significantly violate the expected scientific prior. These sampling artifacts are attributed to the discrepancy between the Gaussian prior and the latent distribution. In Figure~\ref{fig:sampling}(B), we sample from a Gaussian with a smaller variance, $z \sim \mathcal{N}(0.5,0.25)$ and we can see that the quality of samples is better, and more physically consistent. However, this comes at a cost of significantly reduced diversity in the samples as evidenced by the smaller coverage on the 1D scientific prior manifold.

\small{
\subsubsection*{Disclaimer}

 \noindent This document was prepared as an account of work sponsored by an agency of the United States government. Neither the United States government nor Lawrence Livermore National Security, LLC, nor any of their employees makes any warranty, expressed or implied, or assumes any legal liability or responsibility for the accuracy, completeness, or usefulness of any information, apparatus, product, or process disclosed, or represents that its use would not infringe privately owned rights. Reference herein to any specific commercial product, process, or service by trade name, trademark, manufacturer, or otherwise does not necessarily constitute or imply its endorsement, recommendation, or favoring by the United States government or Lawrence Livermore National Security, LLC. The views and opinions of authors expressed herein do not necessarily state or reflect those of the United States government or Lawrence Livermore National Security, LLC, and shall not be used for advertising or product endorsement purposes.}

\bibliographystyle{unsrt}  
\small{
\bibliography{refs}
}
\end{document}